# Surprising dissimilarities in a newly formed pair of 'identical twin' stars

Keivan G. Stassun[1], Robert D. Mathieu[2], Phillip A. Cargile[1], Alicia N. Aarnio[1], Eric Stempels[3], and Aaron Geller[2]

[1]*Department of Physics and Astronomy, Vanderbilt University, Nashville, Tennessee 37235, USA.*

[2]*Department of Astronomy, University of Wisconsin—Madison, Madison, Wisconsin 53706, USA.*

[3]*School of Physics & Astronomy, University of St Andrews, North Haugh, St Andrews KY16 9SS, Scotland*

**The mass and chemical composition of a star are the primary determinants of its basic physical properties—radius, temperature, luminosity—and how those properties evolve with time[1]. Thus, two stars born at the same time, from the same natal material, and with the same mass are 'identical twins,' and as such might be expected to possess identical physical attributes. We have discovered in the Orion Nebula a pair of stellar twins in a newborn binary star system[2]. Each star in the binary has a mass of 0.41±0.01 solar masses, identical to within 2 percent. Here we report that these twin stars have surface temperatures that differ by ~300K (~10%), and luminosities that differ by ~50%, both at high confidence level. Preliminary results indicate that the stars' radii also differ, by 5-10%. These surprising dissimilarities suggest that one of the twins may have been delayed by several hundred thousand years in its formation relative to its sibling. Such a delay could only have been detected in a very young, definitively equal-mass binary system[3] such as that reported here. Our findings reveal cosmic limits on the age synchronisation of young binary stars, often used as tests for the age calibrations of star-formation models[4].**

K.G. Stassun et al.

Astronomers have long relied upon eclipsing binary star systems—in which two stars periodically eclipse one another as they orbit—to measure the basic physical properties of stars and with which to test the most fundamental predictions of theoretical stellar evolution models[1]. Because the two stars in a binary system are presumed to have formed at the same time and from the same parent cloud material, eclipsing binaries permit a direct test of theoretical models with mass as the primary independent variable. In recent years the discovery and analysis of eclipsing binary systems have been fruitfully applied to probe the basic physical properties of newborn low-mass stars and brown dwarfs still in the earliest stages of evolution[5-9].

Par 1802 is a young (age of ~1 million years) eclipsing binary system in the Orion Nebula Cluster recently discovered by us[2], and represents the first known equal-mass eclipsing binary at so young an age. It is thus a particularly sensitive case study with which to test to what extent stars of the same mass, age, and composition are identical, independent of theoretical models. In fact, we find that the components of Par 1802 possess clearly dissimilar surface temperatures and luminosities, and likely dissimilar radii as well.

From radial-velocity measurements of Par 1802 obtained with the Hobby Eberly Telescope's High Resolution Spectrograph, we have previously determined the orbital parameters of the system[2], including a binary mass ratio of $q=1.03\pm0.03$. With the addition of a precise light curve obtained from the 0.9-m telescope at the Kitt Peak National Observatory and the SMARTS telescopes at the Cerro Tololo Inter-American Observatory (Fig. 1), we are now able to perform a combined analysis[10,11] of the light-curve and radial-velocity data, which yields accurate measurements of the fundamental orbital and physical parameters of Par 1802 (Table 1). The refined orbit solution gives an improved mass ratio of $q$=0.98±0.01. Thus, the components of Par 1802 are twins with masses of 0.41±0.01 solar masses (M$_\odot$) in an orbit that is very nearly circular.

The relative depths of the eclipses (Fig. 1) yield the ratio of the stars' surface brightnesses at a wavelength of 0.8 μm, from which we determine the ratio of their surface temperatures to be $T_1/T_2$=1.085±0.007. Our light-curve analysis includes up-to-date model flux spectra for low-mass stars[12], though the simple fact of unequal eclipse depths in a system with a nearly circular orbit by itself makes the finding of unequal surface temperatures incontrovertible.

Par 1802 has previously[13] been assigned a spectral type of M2 with an uncertainty of ±1 subtype, corresponding to a temperature of ~3560±150 K (ref. 4). Similarly, fitting a single-temperature model stellar atmosphere[12] to the observed fluxes of Par 1802 from 0.35μm to 8μm (Supplementary Table 1) gives a temperature of 3800±100 K. This, together with the temperature ratio above, implies component temperatures of $T_1$=3945±15 K and $T_2$=3655±15 K, with the same (correlated) systematic uncertainty of 100 K on both.

In addition, from the observed eclipse durations and orbital velocities we measure the sum of the stars' radii to be $R_1+R_2$=3.51±0.05 solar radii (R$_\odot$). At the precision of our light-curve data (~1%), the nearly circular orbit precludes determining the individual radii from the single-band light curve alone; in a circular system, the primary and secondary eclipse durations are identical, and thus any combination of stellar radii that sum to the same value can reproduce the observed light curve at the level of ~0.1%. Light-curve observations at multiple wavelengths will help to remove this degeneracy in the component radii.

In the meantime, the radii can be estimated from the flux ratio of the two stars. We have performed a spectral decomposition analysis[14,15] of the component spectra of Par 1802 (see Supplementary Information), from which we find $F_2/F_1=0.55\pm0.06$ at 0.6μm. This, combined with the temperature ratio and appropriate bolometric corrections[16], implies a radius ratio of $R_1/R_2=1.08\pm0.05$ or $R_1=1.82\pm0.05 R_\odot$ and $R_2=1.69\pm0.05 R_\odot$. This difference of 5-10% in the radii is consistent with the observed projected rotational velocities (Table 1) if the stars are rotating synchronously at the orbital period with spin axes parallel to the orbital axis. Finally, the ratio of stellar luminosities that results, via the Stefan-Boltzmann law, from the ratios of temperatures and radii is $L_1/L_2=1.58\pm0.10$, which gives $L_1=0.72\pm0.11\ L_\odot$ and $L_2=0.46\pm0.12\ L_\odot$.

We re-fit the observed spectral energy distribution of Par 1802 from 0.35μm to 8μm (Supplementary Table 1) using synthetic spectra[12] of cool stars with the above temperatures and radii (Fig. 2). Only the distance and reddening due to extinction were varied in the fit. The observed spectral energy distribution is very well fit with an extinction of $A_V=0.5\pm0.2$ visual magnitudes and a distance of 420±15 parsecs, consistent with the low-mass stellar population associated with the Orion Nebula at 480±80 parsecs[17]. This star-forming region is very young, with an estimated age of $1^{+2}_{-1}$ million years[13].

There is weak evidence for an excess of infrared emission at wavelengths longer than 5 μm (Fig. 2), perhaps indicative of a circumbinary disc and/or a faint tertiary body. In this light, the slight eccentricity of the orbit (Table 1) might also be taken as evidence for a tertiary body in the system. Flux measurements at longer wavelengths will be required to verify this possibility.

Unequal temperatures and luminosities for the equal-mass stars of Par 1802 are securely established in the current analysis, and unequal radii are also suggested. Interestingly, some stellar evolution models for young low-mass stars[3] predict that stars with masses of 0.4 M$_\odot$ undergo a brief period of rapid evolution at an age of ~1 million years (Fig. 3). In particular, the observed temperatures, luminosities, and radii of Par 1802 are consistent with the model predictions if the warmer, larger, more luminous star is interpreted as being slightly less evolved than its companion. Because of the predicted rapidity of evolution, a 'lag' of only a few hundred thousand years would be required (Fig. 3).

Such an age difference is only observationally detectable in an equal-mass binary system during the first few million years of its evolution, where the models predict that changes in the stars' physical properties are fastest and most pronounced. At later times any physical signs of non-synchronisation will have become smaller than 1%, below the precision limit of the best observations (Fig. 3). On the other hand, in a young binary with unequal-mass components, uncertain theoretical models will convert precise physical differences into an imprecise age difference. That Par 1802—the first very young, definitively equal-mass binary to be studied—shows evidence for age differences between its stars may suggest that this is a common feature of young binaries.

Current theories of binary formation are largely silent on the formation of such short-period binaries, and so we have no theoretical context within which to interpret such a difference in evolutionary age. Certainly the lack of synchronisation of the stellar evolution clocks will provide a new clue into the formation processes of short-period binaries.

Alternatively, the stars may be the same age but have differing properties despite their very similar masses and (presumably) composition. There is mounting evidence that strong magnetic fields on the surfaces of young stars may alter their physical properties[8,18,19,20]. Thus, for example, one of the stars in Par 1802 may possess a strong magnetic field that is moreover substantially different in strength or geometry compared to its companion. However, we have observed that the strength of the Balmer α line of hydrogen, a commonly used tracer of magnetic activity in low-mass stars, is very weak in both components of Par 1802 (at most a few hundred milli-ångströms of emission)[7]. Moreover, the observational and theoretical evidence to date for the effects of magnetic fields on stellar properties indicates that only the surface temperatures and radii of the stars should be affected. The luminosities of the stars are not predicted to be significantly modified by magnetic fields because the luminosities are primarily determined by internal processes[19]. These expectations are at odds with the finding of a factor of ~ 2 luminosity difference between the stars in Par 1802. In any case, even if it were correct, this hypothesis would still beg the question of why these two stars of essentially identical masses and rotation rates do not possess similar magnetic properties.

Finally, young binary systems have been used as testbeds of theories of early stellar evolution. The agreement of theoretical ages derived for each star in a binary system is taken as a self-consistency test for pre-main-sequence stellar evolution models[4]. The lack of synchronisation in Par 1802 suggests a precision limit of several hundred thousand years for such empirical tests.

Most importantly, Par 1802 provides the first direct evidence that birth order in 'identical twin' stars can manifest itself as observable physical differences between the two stars—at least when they are very young.

We are grateful to A. Prsa for software used in our analyses. This work is supported by grants to K.G.S. and R.D.M. from the National Science Foundation, and a Cottrell Scholar award to K.G.S. from the Research Corporation. K.G.S. acknowledges the hospitality of the Space Telescope Science Institute's Caroline Herschel Distinguished Visitor program.

Correspondence should be addressed to K.G.S. (e-mail: keivan.stassun@vanderbilt.edu).

**Supplementary Information** is linked to the online version of the paper at www.nature.com/nature.

K.G. Stassun et al.

**Table 1: Orbital and physical parameters of Par 1802**

| | |
|---|---|
| Period, $P$ | 4.673843±0.000068 days |
| Time of periastron passage | 2003.834996±0.000055 |
| Eccentricity, $e$ | 0.029±0.005 |
| Orientation of periastron, $\omega$ | 266.1±1.8° |
| Semi-major axis, $a \sin i$ | 0.0501±0.0006 AU |
| Centre-of-mass velocity, $\gamma$ | 23.7±0.5 km s$^{-1}$ |
| Mass ratio, $q \equiv M_2/M_1$ | 0.98±0.01 |
| Total mass, $(M_1+M_2)\sin^3 i$ | 0.768±0.028$M_\odot$ |
| Inclination, $i$ | 78.1±0.6° |
| Primary mass, $M_1$ | 0.414±0.015$M_\odot$ |
| Secondary mass, $M_2$ | 0.406±0.014$M_\odot$ |
| Sum of radii, $R_1+R_2$ | 3.51±0.05$R_\odot$ |
| Surface temperature ratio, $T_1/T_2$ | 1.084±0.007 |
| Primary rotation speed, $v_1 \sin i$ | 17±2 km s$^{-1}$ |
| Secondary rotation speed, $v_2 \sin i$ | 14±3 km s$^{-1}$ |

AU, astronomical units.

We simultaneously fit the radial-velocity data from Cargile *et al*. (ref. 2) and the light curve newly presented here (Fig. 1) using a standard detached-eclipsing-binary model[10,11]. The code assumes full Roche geometry according to the formalism of Kopal[10], and includes model atmospheres[12] to determine intensities over the stellar discs. We adopted a linear limb darkening law and allowed the code to calculate reflection effects, adopting a bolometric albedo of 0.5, typical for fully convective stellar atmospheres. In order to maintain control of the solution and its many parameters, we performed this fitting in stages. In the initial stage we fixed the orbital parameters at the values from Cargile *et al.* (ref. 2), and assumed an average surface temperature of 3800K for the components (see text) and solar metallicity. This allowed us to obtain initial estimates of the component temperatures and the sum of the radii, and the system inclination. With these initial values so determined, we then iteratively improved the solution by first fitting the eccentricity and orientation of periastron, and then performing a final fit in which all of the orbital and component parameters of the system were fit freely. Uncertainties in the parameters represent standard $1\sigma$ (s.e.m.) formal errors from the covariance matrix of the eclipsing-binary model fit[10,11]. The most important degeneracy in the solution is between the component radii; due to the nearly circular orbit, virtually any combination of radii that sum to the same total value will equally well fit the light curve. We have measured the projected rotation speeds, *v* sin*i*, of the two components by comparing the widths of the observed cross-correlation peaks[2] to those of an M2 star whose spectrum was rotationally broadened artificially. The *v* sin*i* values and uncertainties are the averages and standard deviations resulting from five observations of Par 1802 near maximum radial-velocity separation of the two components[2]. Note that both here and in Cargile *et al.* (ref. 2) the more luminous star (here the formally more massive star) is identified as the 'primary.'

K.G. Stassun et al.

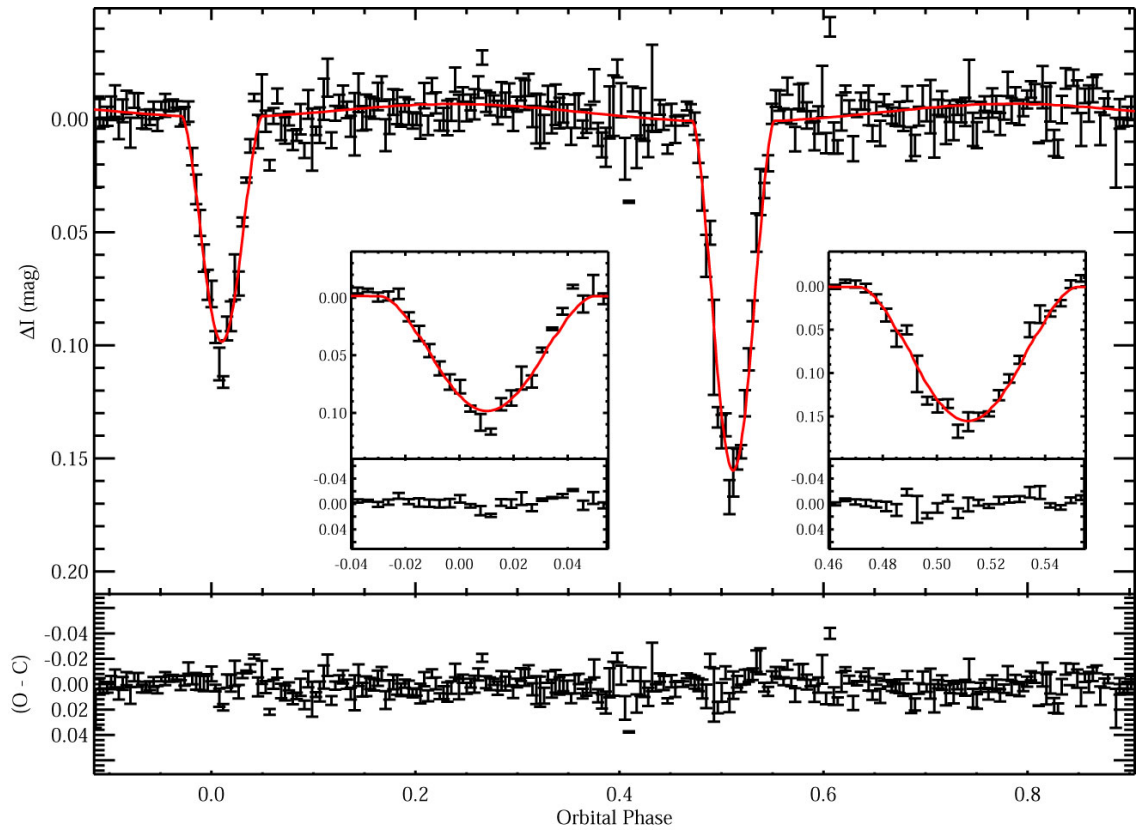

**Figure 1: Light curve of Par 1802 at I band (0.8 μm).** We repeatedly imaged Par 1802 with the 0.9-m telescope at the Kitt Peak National Observatory, and the SMARTS 0.9-m, 1.0-m, and 1.3-m telescopes at the Cerro Tololo Inter-American Observatory, from December 1994 to March 2007. In total, 2,209 flux measurements were obtained on 418 separate nights and with an average cadence of 5–6 measurements per night. The typical relative uncertainty on the individual flux measurements is ~1 percent. A phase dispersion minimization (PDM) analysis[21] reveals an unambiguous period of P=4.673843±0.000068 days. The measurements are shown here as differential magnitudes ΔI, folded on the above period and phased relative to periastron passage (that is, closest approach of the two stars to one another), as determined from the orbit solution (Table 1). The measurements have been re-sampled into 250 bins equally spaced in phase;

each data point plotted represents the average of ~9 measurements, and error bars represent the r.m.s. of the values that were averaged together. Note that the fitting procedure (see Table 1) was performed on the raw data, not the re-sampled light curve shown here for visual clarity. The ratio of eclipse depths provides a direct measure of the ratio of surface temperatures, with the deeper eclipse corresponding to the eclipse of the hotter component. Due to the eccentricity and orientation of periastron, the eclipse of the primary component occurs near orbital phase 0.52. A model fit to the light curve incorporating the orbital and physical parameters of Par 1802 (Table 1) is also shown (solid red curve). Insets show the eclipses in detail, while the lower panel shows the residuals in magnitudes of the data relative to the model (observed minus calculated).

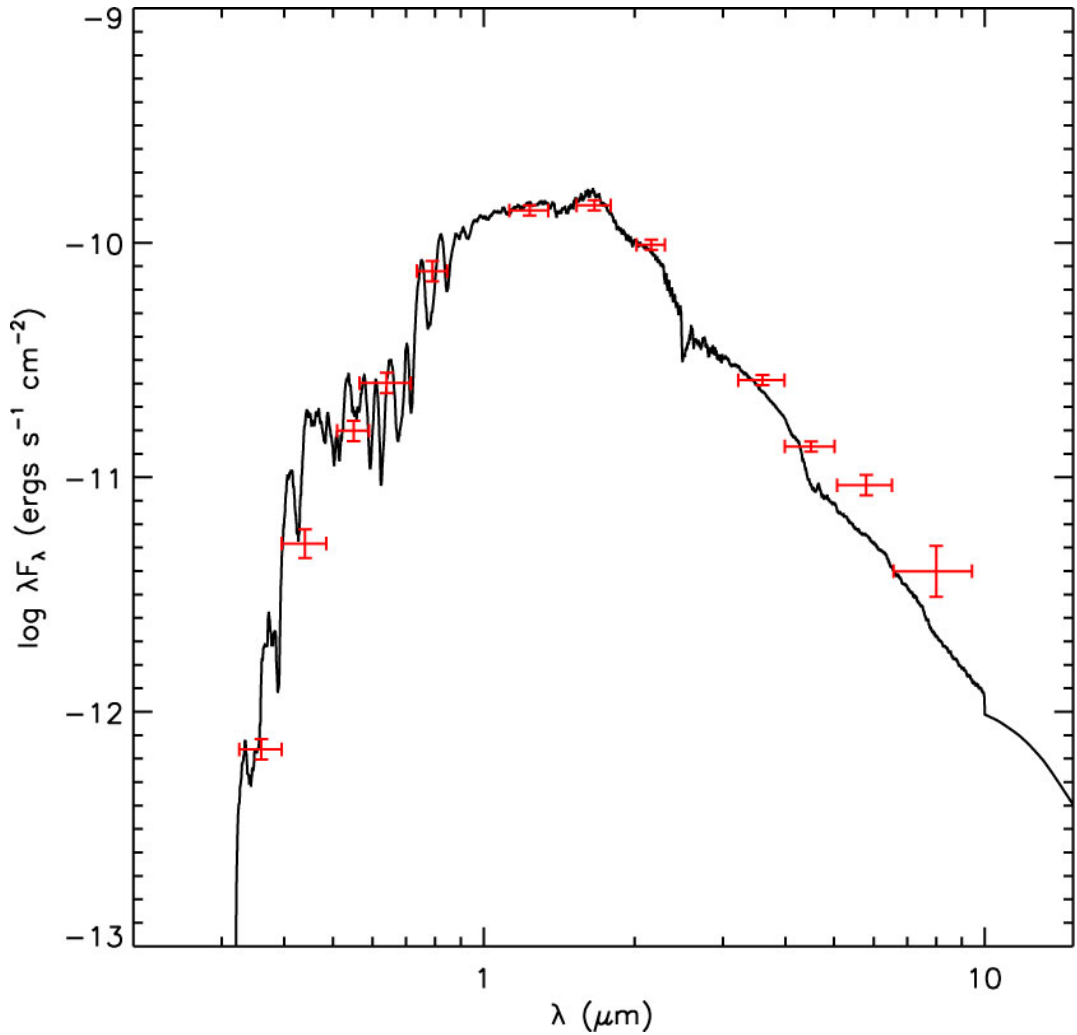

**Figure 2: Spectral energy distribution of Par 1802.** Broad-band flux measurements of Par 1802 from 0.35μm to 8μm (Supplementary Table 1; Refs. 22-30) are shown in red. Vertical error bars represent s.e.m. uncertainties on the flux measurements, horizontal bars represent the filter bandpasses used for the flux measurements. The solid curve is a composite of two synthetic spectra[12] of young, low-mass stars with temperatures, masses, and radii corresponding to those measured for the components of Par 1802 (Table 1). Fitting for extinction and distance, we find $A_V$=0.5±0.2 magnitudes and $d$=420±15 parsecs.

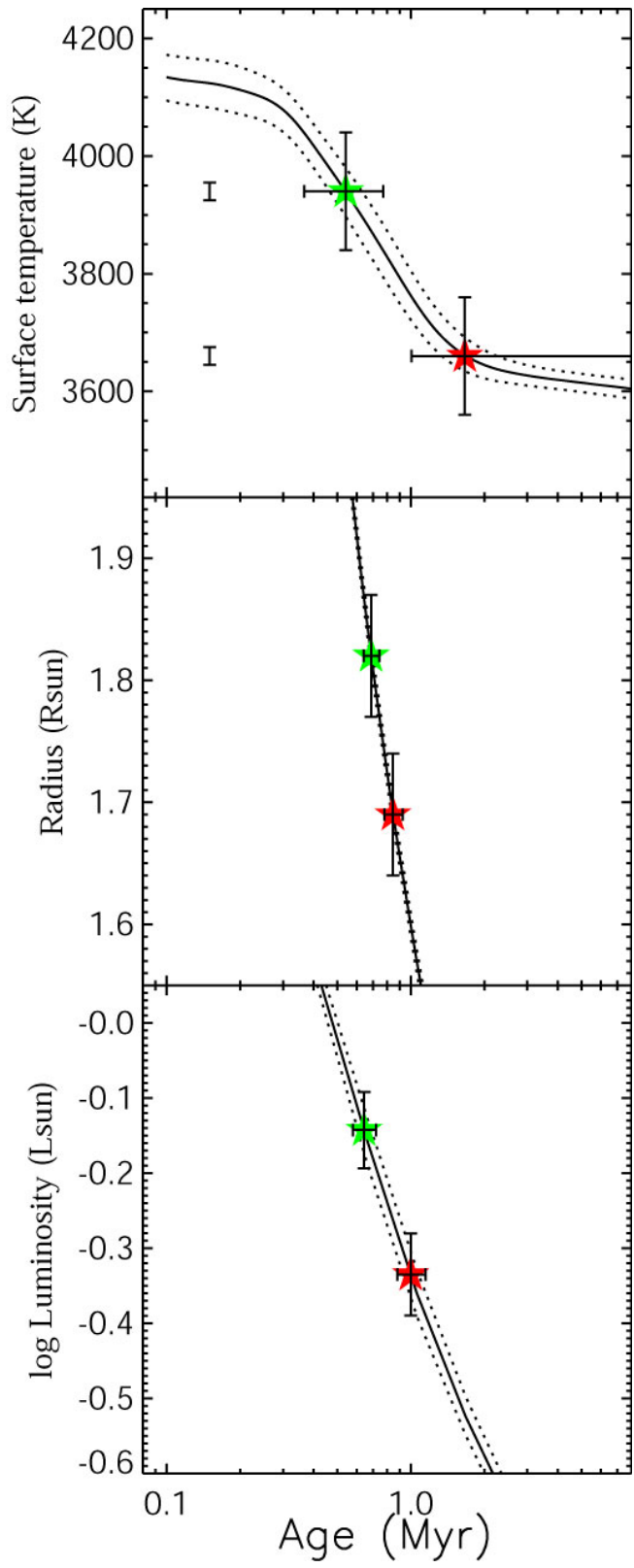

**Figure 3: Comparison of physical properties of Par 1802 to theoretical predictions.** In each panel, the solid line shows the predicted evolution of a 0.41$M_\odot$ star with solar composition from the theoretical models of D'Antona & Mazzitelli (ref. 3). Dotted lines show the result of changing the stellar mass by ±0.015 $M_\odot$, representative of the uncertainties on the measured masses of Par 1802 (Table 1). The measured properties of the primary and secondary components of Par 1802 are shown as green and red symbols, respectively. In the top panel, the small vertical bars at left represent the measurement errors alone (i.e. not including the ~100K systematic uncertainty on the absolute temperature scale) resulting from the precisely measured temperature ratio. Horizontal error bars represent the range of ages for which the theoretical models are consistent with the measurements within the uncertainties (including systematic uncertainties). The stellar luminosities plotted in the bottom panel are calculated from the measured radii and surface temperatures. Note that the uncertainties on the temperatures, radii, and luminosities are not independent between the two stars, as they are connected by precisely determined ratios; thus, e.g., the primary star cannot be forced cooler while simultaneously forcing the secondary warmer. The nominal age of the Orion Nebula Cluster is ~1 million years[13].

K.G. Stassun et al.

## Supplementary Information

**Measurement of the flux ratio of the components of Par 1802**

To determine the flux ratio of the two components of Par 1802, we performed an analysis of the observed spectra from Cargile *et al.* (ref. 2) using two techniques: tomographic reconstruction and two-dimensional cross-correlation.

First, we reconstructed the individual spectra of the primary and secondary component using tomographic reconstruction[14], a standard technique for the analysis of spectroscopic binaries. This technique uses a gradient-search algorithm to determine the primary and secondary spectra that, when recombined, minimize the $\chi^2$ of the residuals with respect to the observed spectra. If observations of at least two orbital phases are available, this provides enough information to successfully reconstruct the component spectra. With the relative velocities of the two components known *a priori*, the only free parameter in this algorithm is the flux ratio of the two components. This parameter regulates the relative contribution of the primary with respect to the secondary by scaling their continuum fluxes. Since absorption line depths are measured with respect to the continuum level, changing the flux ratio effectively determines the ratio of overall spectral line depth between the two component spectra. Since the overall line depth is a diagnostic for stellar temperature, and we also have good estimates of the effective temperatures of the components from the light-curve and spectral energy distribution analyses (Figs. 1 and 2), we can compare our reconstructed spectra with observed template spectra of stars with known temperatures. This allows us to determine the flux ratio that best reproduces the expected spectra of the primary and secondary. We have applied this technique using three observations from Cargile *et al.* (ref. 2) in which the components are well separated in velocity but at different orbital phases, and have compared the reconstructed spectra (Supplementary Figure 1) with spectral templates of M1V for the primary and M3V star for the secondary. We used three spectral orders in the region around 0.6μm. From this analysis we find the flux ratio of the two components to be $F_2/F_1$=0.50±0.07, where the uncertainty is estimated from the scatter in the multiple observations and orders used.

We also examined the flux ratio using a two-dimensional cross-correlation (TODCOR) analysis[15]. In this case the flux ratio is determined by simultaneously cross-correlating a spectrum of the binary against two spectral templates and finding the flux ratio that produces the strongest two-dimensional correlation. The TODCOR analysis does not "reconstruct" the observed spectrum using information from spectra at multiple orbital phases as does the tomographic reconstruction analysis. However it does have the advantage of allowing different combinations of spectral templates to be tried and in principle can provide an independent determination of the spectral types of the two components. We applied the TODCOR analysis to the observation of Par 1802 from Cargile *et al.* (ref. 2) in which the two components are at maximum velocity separation (see Fig. 2 in that paper), again using three spectral orders in the region around 0.6μm. We used all possible combinations of spectral templates with spectral types of K2, K3, K5, K7, M1, and M2 observed by Cargile *et al.* (ref. 2). The analysis gives a mild preference for spectral types of K7 and M2, but virtually any combination of templates in the late-K to early-M spectral range yields very similar results, likely because the ~300 K difference in temperature between the components does not produce strong differences in the spectral features of late-K to early-M dwarfs. From this analysis we find the flux ratio of the two components of Par 1802 to be $F_2/F_1$=0.61±0.10, where the uncertainty is estimated from the scatter in the multiple orders used.

The flux ratios determined via the two analyses are compatible within the uncertainties. For our present purposes, we adopt the mean of the two determinations and the uncertainty on the mean, i.e. $F_2/F_1 = 0.55 \pm 0.06$.

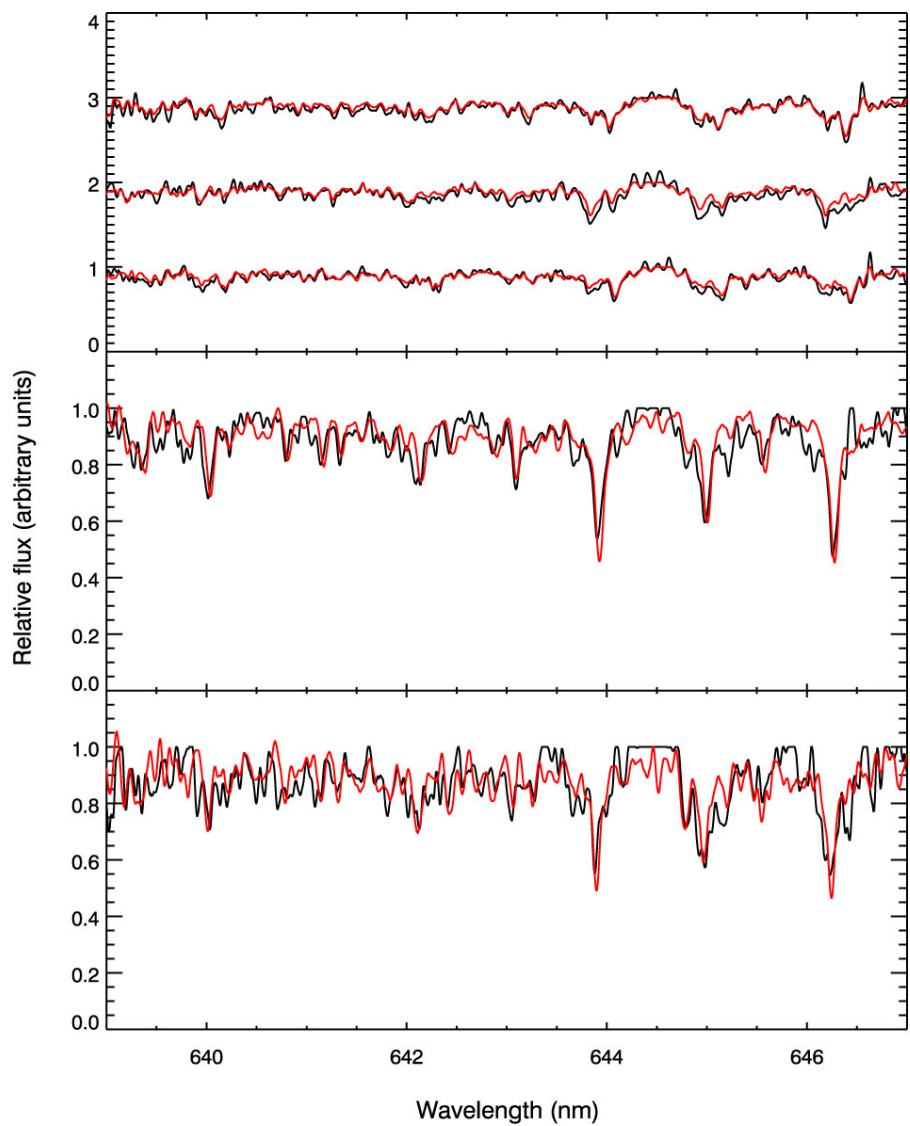

**Supplementary Figure 1: Tomographic reconstruction analysis of the flux ratio in Par 1802.** Three observations of Par 1802 from Cargile *et al.* (ref. 2) were analysed via tomographic spectral reconstruction[14] to determine the spectroscopic flux ratio of the two components. Shown in the top panel are portions of the three

observed spectra near 0.6μm (black) with the tomographically reconstructed spectra overlaid (red). The reconstructed primary (middle) and secondary (bottom) spectra (black) are compared to standard star spectra (red). A secondary-to-primary flux ratio of $F_2/F_1$=0.50±0.07 best reproduces the observed spectra.

**Broadband flux measurements of Par 1802**

**Supplementary Table 1: Broadband flux measurements of Par 1802**

| Bandpass | Wavelength, $\lambda$ ($\mu$m) | Flux, $F_\lambda$ (mJy) | Source | Ref. |
| --- | --- | --- | --- | --- |
| U | 0.36 | 0.083±0.0083 | ESO WFI | 22 |
| B | 0.45 | 0.769±0.108 | ESO WFI | 22 |
| V | 0.55 | 2.894±0.2894 | ESO WFI | 22 |
| $R_C$ | 0.67 | 5.388±0.5388 | NOMAD | 23 |
| $I_C$ | 0.80 | 19.96±1.996 | ESO WFI | 22 |
| J | 1.2 | 56.6±2.83 | 2MASS | 24 |
| H | 1.5 | 80.1±4.01 | 2MASS | 24 |
| $K_S$ | 2.2 | 70.6±3.53 | 2MASS | 24 |
| IRAC1 | 3.6 | 31.2±1.56 | *Spitzer* | 30 |
| IRAC2 | 4.5 | 20.3±1.02 | *Spitzer* | 30 |
| IRAC3 | 5.8 | 17.9±1.79 | *Spitzer* | 30 |
| IRAC4 | 8.0 | 10.6±2.65 | *Spitzer* | 30 |

mJy, milli-Janskys.

The broadband fluxes of Par 1802 (Fig. 2) were compiled from several sources. *U*, *B*, *V*, and *I*$_C$ magnitudes were obtained from an HST Orion Treasury project compilation[22] including European Southern Observatory (ESO) wide-field imager (WFI) photometric data. The Naval Observatory Merged Astrometric Dataset (NOMAD) catalogue[23] provided the $R$ band measurement, while near infrared $J$, $H$, and $K_S$ magnitudes were taken from the 2-Micron All Sky Survey (2MASS) catalogue[24]. These magnitudes were then converted to fluxes utilising standardized zero-points[25-29]. At mid-infrared wavelengths, images were obtained from *Spitzer Space Telescope* Infrared Array Camera (IRAC) archival data[30]. To measure IRAC fluxes, the IRAF task *apphot* was used to perform aperture photometry on post-baseline calibrated images from the *Spitzer* data-processing pipeline. The source and sky subtraction annuli were chosen to include the full source extent and remove sky background. After applying appropriate aperture corrections, a calibrated transformation from the *Spitzer* Data Analysis Guide was used to convert integrated counts to flux. Uncertainties above reflect adopted fractional uncertainties of 5-25% to account for likely systematic errors including typical levels of source variability at a given wavelength.